\documentclass[fleqn,10pt]{wlscirep}
\usepackage[utf8]{inputenc}
\usepackage[T1]{fontenc}
\title{Heatmaps in soccer: event vs tracking datasets}

\author[1,2]{D. Garrido}
\author[1,3]{Borja Burriel}
\author[4]{R. Resta}
\author[4]{R. L\'opez del Campo}
\author[1,2,5,*]{J.M. Buld\'u}
\affil[1]{Complex Systems Group \& GISC, Universidad Rey Juan Carlos, 28933 Madrid, Spain}
\affil[2]{Laboratory of Biological Networks, Center for Biomedical Technology, UPM, Pozuelo de Alarc\'{o}n, 28223 Madrid, Spain}
\affil[3]{Polytechnical University of Catalonia, 08034 Barcelona, Spain}
\affil[4]{Mediacoach - LaLiga, 28043 Madrid, Spain}
\affil[5]{Unmanned Systems Research Institute, Northwestern Polytechnical University, Xi'an 710072, China}

\affil[*]{email: javier.buldu@urjc.es}


\begin{abstract}
We investigate how similar heatmaps of soccer players are when constructed from (i) event datasets and (ii) tracking datasets. When using event datasets, we show that the scale at which the events are grouped strongly influences the correlation with the tracking heatmaps. Furthermore, there is an optimal scale at which the correlation between event and tracking heatmaps is the highest. However, even at the optimal scale, correlations between both approaches are moderate. Furthermore, there is high heterogeneity in the players' correlation, ranging from negative values to correlations close to the unity. We show that the number of events performed by a player does not crucially determine the level of correlation between both heatmaps. Finally, we  analyzed the influence of the player position, showing that defenders are the players with the highest correlations while forwards have the lowest.
\end{abstract}
\begin{document}

\flushbottom
\maketitle
% * <john.hammersley@gmail.com> 2015-02-09T12:07:31.197Z:
%
%  Click the title above to edit the author information and abstract
%
\thispagestyle{empty}

%\noindent Please note: Abbreviations should be introduced at the first mention in the main text – no abbreviations lists. Suggested structure of main text (not enforced) is provided below.

\section*{Introduction}

The recent ability to capture player positions through optical tracking has made heatmaps to become a standard tool of analysis in soccer. From the technical staff of a team to journalist and fans, a diversity of people uses and analyses players' heatmaps, with the aim of understanding what parts of the pitch were occupied by a given player during a match. This is valuable information for technical staffs not only to prepare matches by analyzing the expected position of rivals but to understand to what extent the tactical roadmap was followed in the after-match analysis. However, although the use of heatmaps in soccer is quite common, the analysis of their fundamental properties is still under development. In that sense, Clemente et al.
 \cite{clemente(2013)} analysed the difference between heatmaps of a team with and without ball possession. The authors showed that it was possible to interpret a match according to the variation of the occupation of the field in these two phases. More recently, Moura et al.  \cite{moura(2017)}  focused on the variability of the player's position during a whole match. They analysed the tracking datasets of players during the 2012 UEFA European Championship, quantifying the main directions of the variability of each player (and team) and showing that external midfielders were the players with the highest variability. Another interesting application  of the analysis of players' heatmaps is the development of algorithms to detect team's formations \cite{bialkowski(2014),machado(2017)}. 

However, there is a crucial point in the construction of heatmaps: the datasets. There are two main ways of constructing heatmaps (i) using event datasets or (ii) using tracking datasets.  In the former, all actions performed by a player during a match are tagged with their corresponding Euclidean position and, next, they are taken as the input to infer where a player has been placed during the match. On the other hand, tracking datasets contain the actual position of a player with a resolution of some centimetres and sampling rates below a second  \cite{gudmundsson(2017)}.

In this paper we analyze the relation between players' heatmaps generated from (i) event datasets and (ii) tracking datasets. The motivation behind this analysis comes from the fact that, during the last years, the use of event-based heatmaps has become widely used in the context of sports media and soccer analysts. In this way, it is common to see players' heatmaps in the majority of soccer specialised media as well as in a diversity of platforms related to the analysis of soccer matches. However, less attention is paid to the origin of these heatmaps, remaining unclear in many cases. The fact that tracking datasets are more costly and, sometimes, inaccessible makes event datasets the source generating  most of such heatmaps. These datasets, which also have a certain cost (depending on the provider), consist of all actions performed by any player during the whole match, including passes, shots, faults, ball touches or air duels, to name a few. 
With these datasets, it is possible to construct a player's heatmap that, adequately filtered, seems to reproduce the position of a player during the whole match. However, when the origin of the datasets (i.e., events) is forgotten, there is space for misunderstandings. For example, this is the case of Luis Suarez's heatmap obtained during the match between F.C. Barcelona and Bayern Munich at the quarter-finals of the 2020 Champions League. Luis Suárez made 24 passes during the 90 minutes. Nine of them were from kick-offs, leading to a heatmap with a maximum at the centre of the pitch \cite{suarez}. To clarify to what extent event and tracking heatmaps are similar, we computed the correlation between them at several scales, analysing the influence of the number of events and the role of the player position in the average correlation between both kinds of heatmaps.

\section*{Methods}

\subsection*{Datasets}

As explained in the Introduction, we used two kinds of datasets, namely, event and tracking datasets. On the one hand, tracking datasets have been supplied by LaLiga software Mediacoach  \cite{mediacoach}. The tracking datasets consist of the position of $N=1320$ match observations of elite soccer players during $L=60$ matches of the Spanish first and second divisions, LaLiga Santander ($L_1=30$) and LaLiga Smartbank ($L_2=30$), respectively. A multi-camera tracking system recorded each player’s position on the pitch with a sampling rate of $\Delta f$=25 frames/second  \cite{linke(2020)}. Each multi-camera unit contained three cameras with a resolution of 1920x1080 pixels that were synchronized to provide a panoramic picture and created the stereoscopic view for triangulating the players and the ball. An experienced operator corrected the position of players in the case of a temporal loss of any location. Datasets obtained by the Mediacoach system have been previously validated with GPS  \cite{felipe(2019),pons(2019)}.
Each player was tagged with his position in the team, which has been split into four categories: goalkeepers ($N_1=120$), defenders ($N_2=471$) and midfielders ($N_3=520$), forwards ($N_4=209$). Note that $N_i$ refers to the total number of participations tagged with the corresponding position. Since datasets contained several fixtures, the same player could have contributed more than one time to $N_i$. Next, we divided the pitch into $M \times M$ divisions and developed an algorithm to evaluate the time accumulated by each player inside each of the $P= M \times M$ regions of the pitch. As we will see later, the number of pitch divisions ($P$) will be one of our control parameters.

On the other hand, the event datasets were supplied by Opta  \cite{opta} and contained the 2-dimensional Euclidean position of all actions performed by players during the same set of matches (see Ref.  \cite{optadescription} for a full description of all recorded events).
 
\subsection*{Event vs Tracking heatmaps}

Figure 1 shows an example of the event (left plots) and tracking (right plots) heatmaps of two different players, a central defender (top plots) and a forward (bottom plots). In both cases, event and tracking heatmaps are different, however, in the case of the defender, there is a high overlap between them. When looking at the forward player, we can observe that the discrepancy is high. The fact that most of his actions are quite dispersed over the pitch leads to a poor agreement between both kinds of heatmaps.

 %%%%%%%%%%%%%%%%%%% FIGURE 01
\begin{figure}[h]
\centering
\includegraphics[width=6.5 cm]{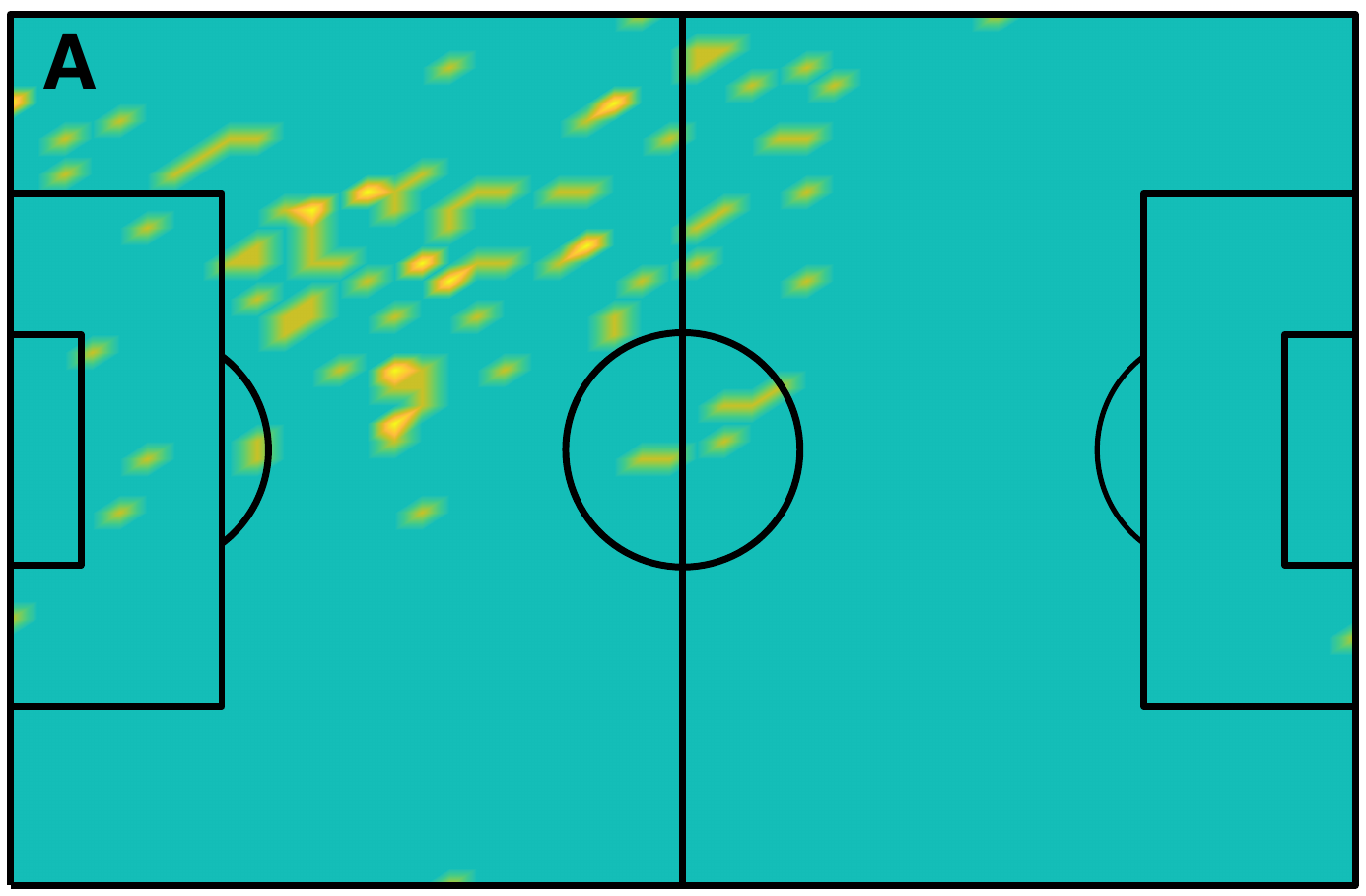}
\includegraphics[width=6.5 cm]{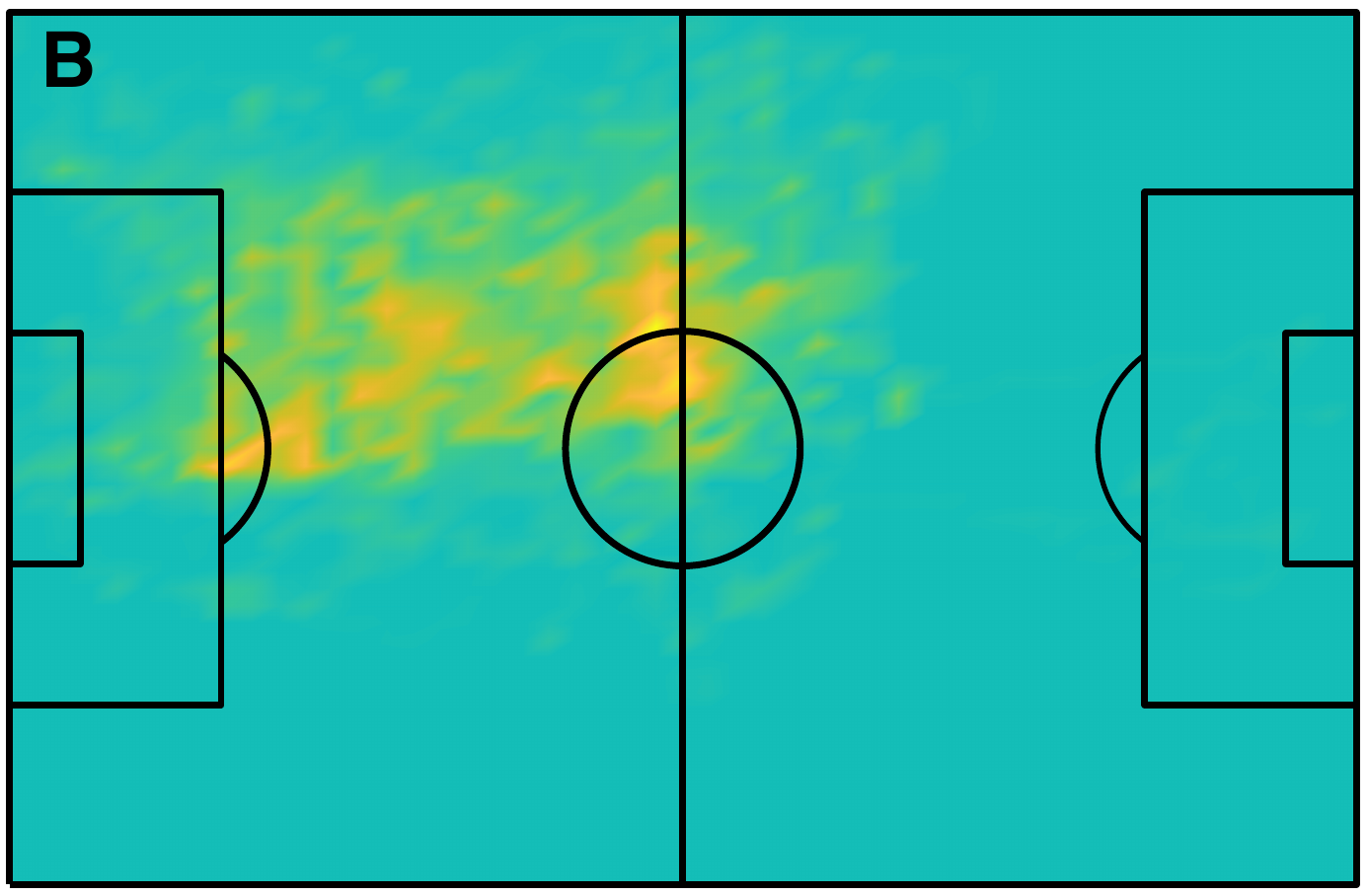}
\includegraphics[width=6.5 cm]{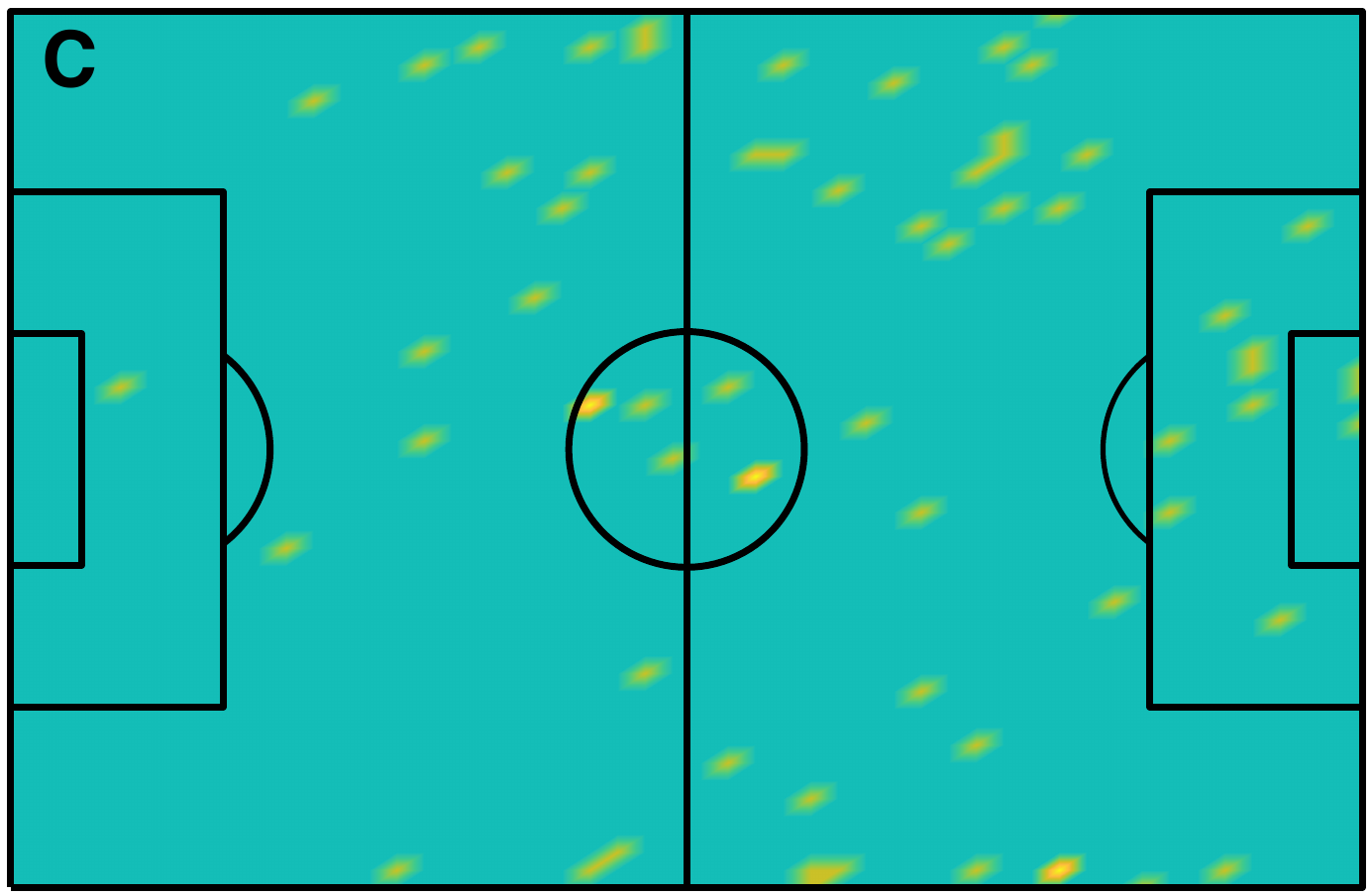}
\includegraphics[width=6.5 cm]{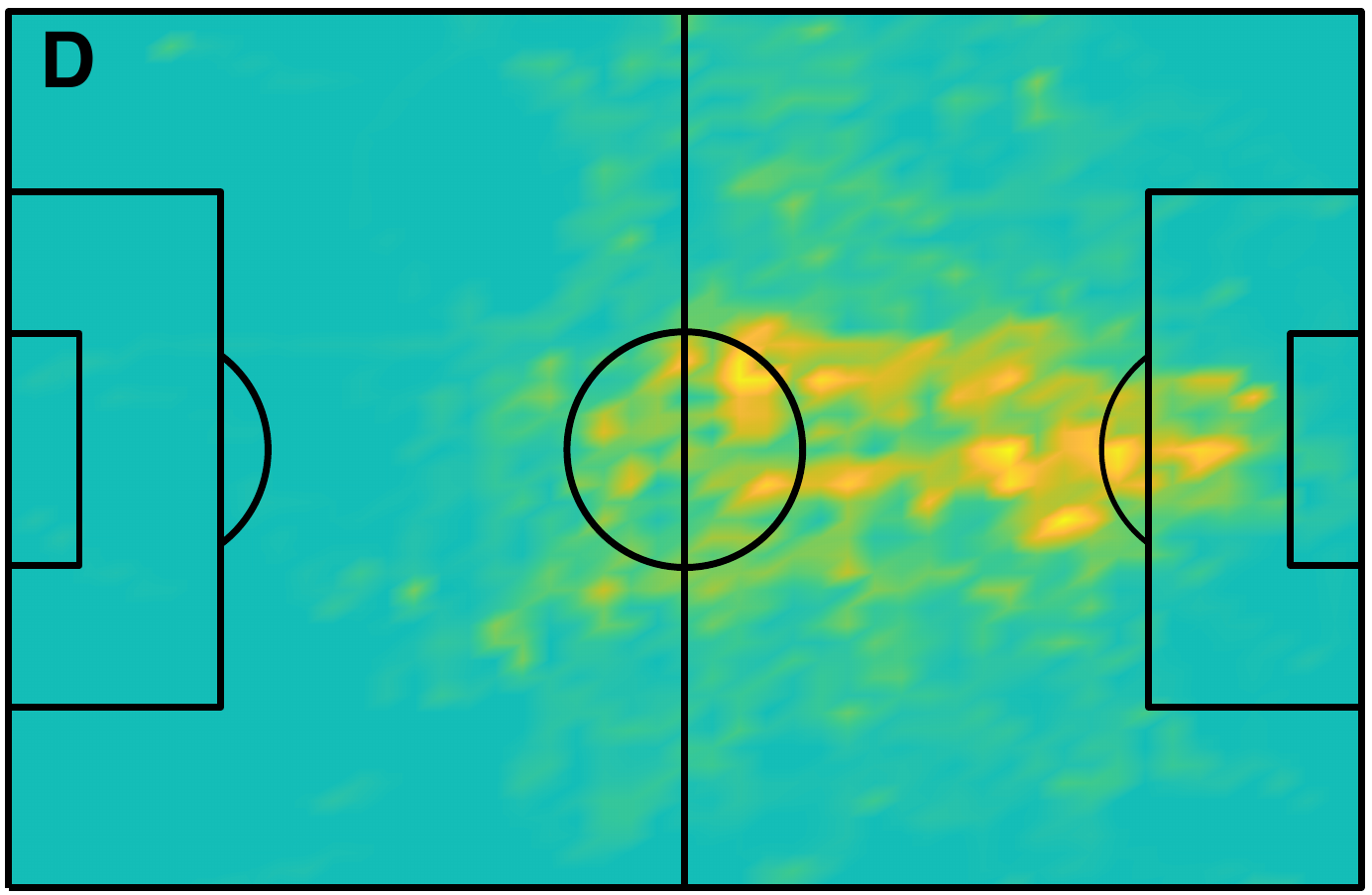}
\caption{Event vs tracking datasets. Plots A and B correspond, respectively, to the heatmaps obtained from the event (A) and tracking (B) datasets for a given player, in this case, a central defender. Plots C and D correspond to a forward player of the same team during the same match. Note that, while in the upper plots (defender), events and positional tracking cover similar regions, bottom plots are quite different due to the dispersion of the events performed by the forward player. \label{f01}}
\end{figure}
 %%%%%%%%%%%%%%%%%%% FIGURE 01

\begin{figure}[t]
\centering
\includegraphics[width=6.5 cm]{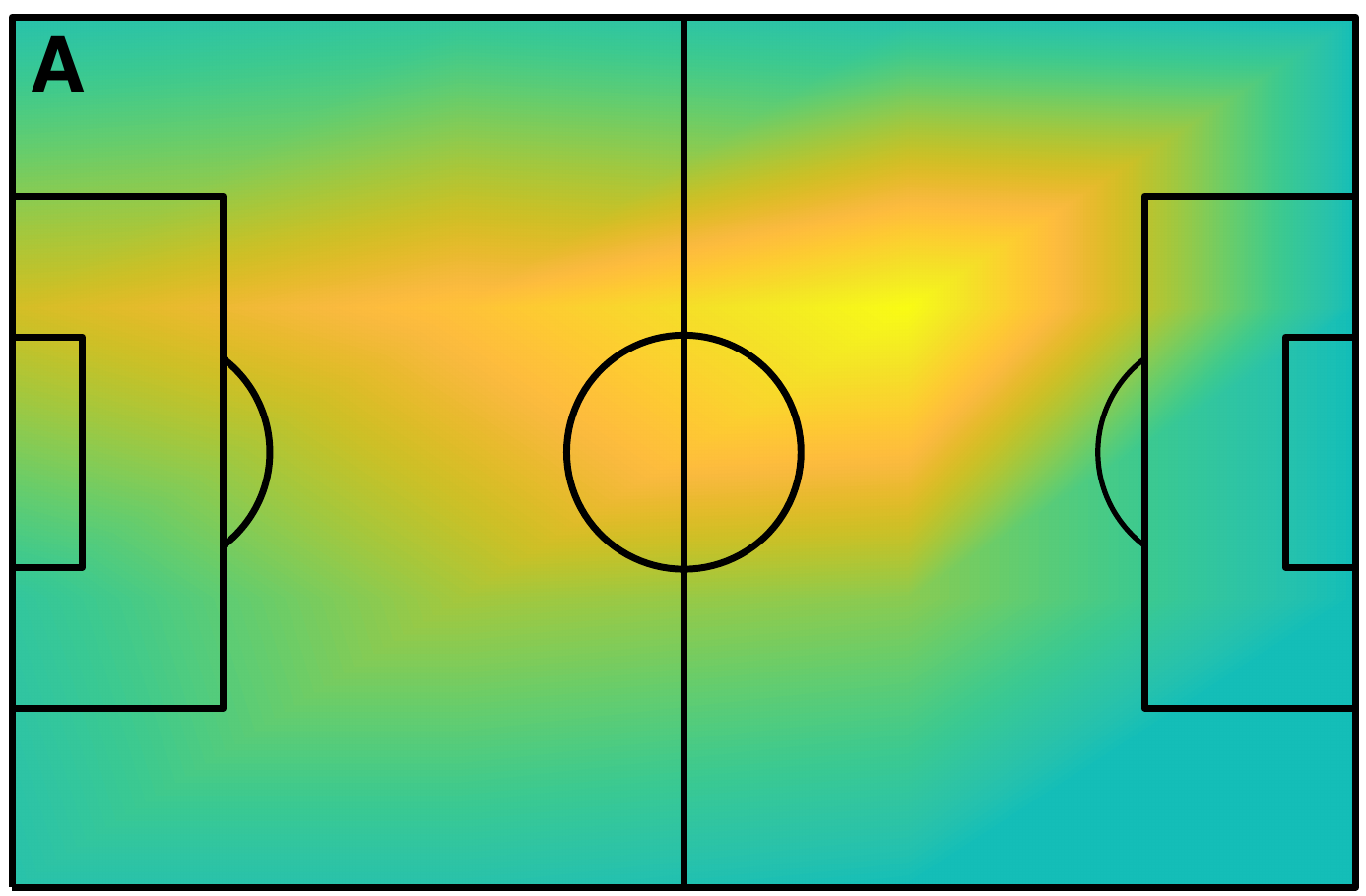}
\includegraphics[width=6.5 cm]{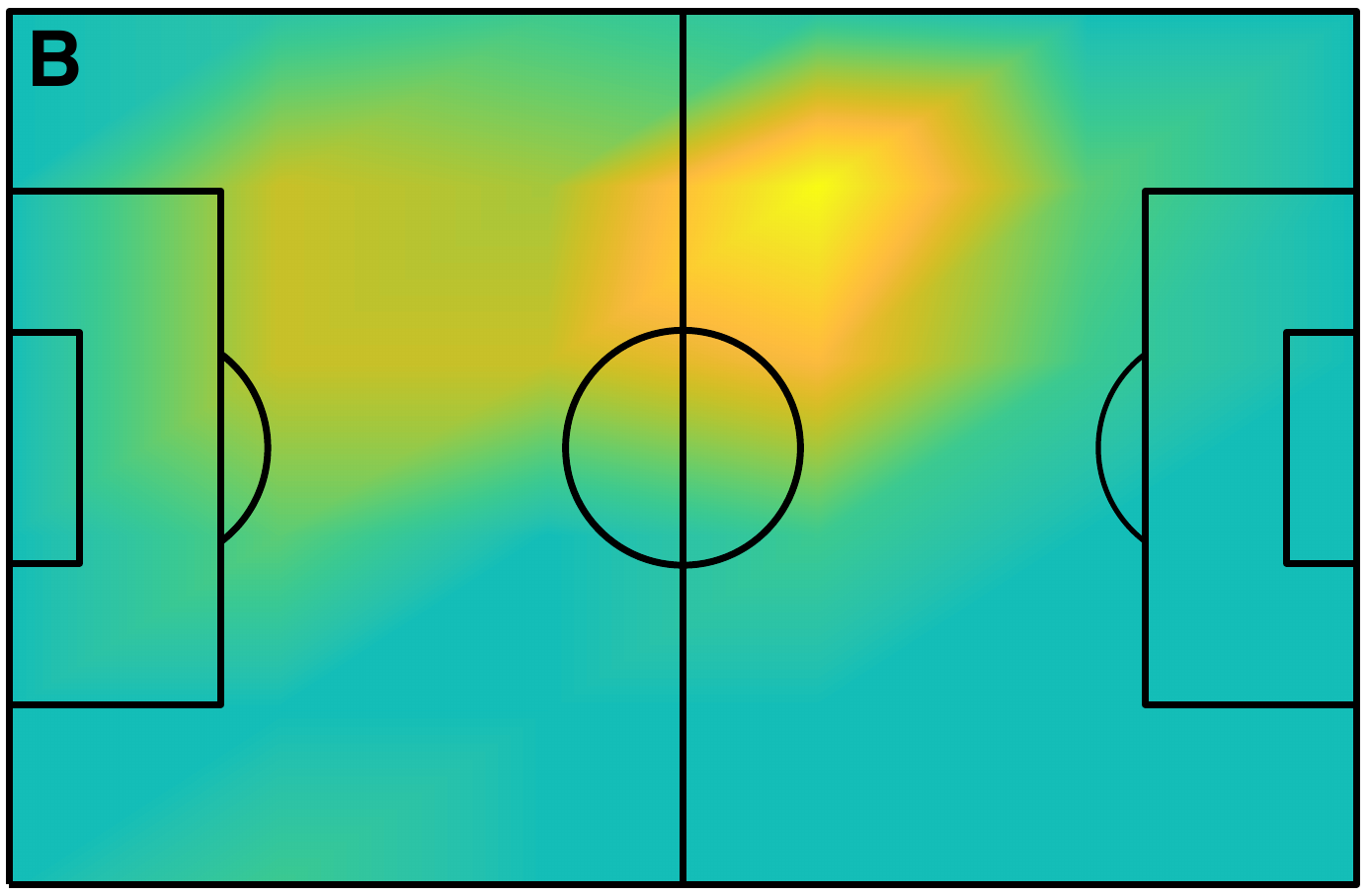}
\includegraphics[width=6.5 cm]{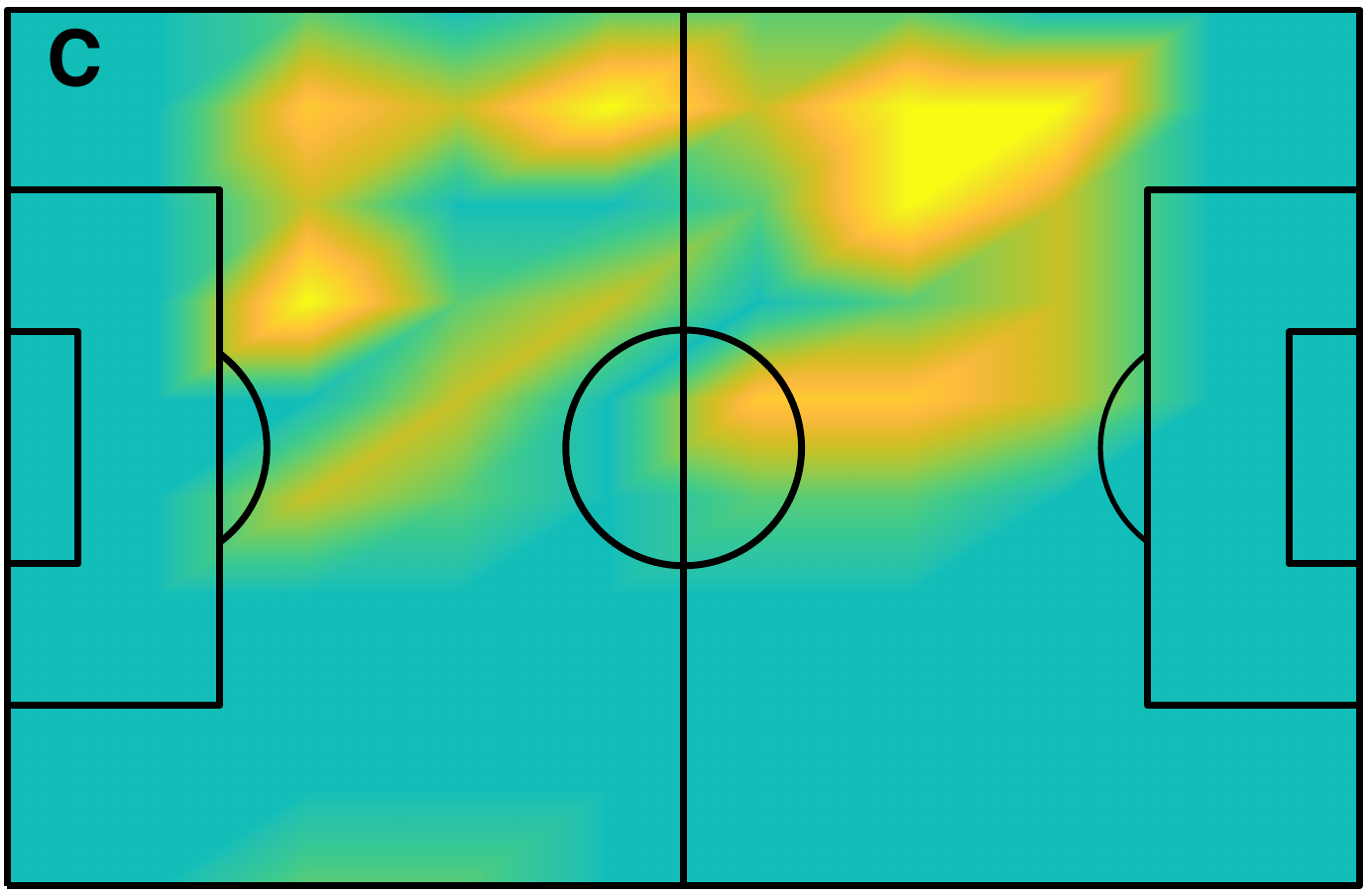}
\includegraphics[width=6.5 cm]{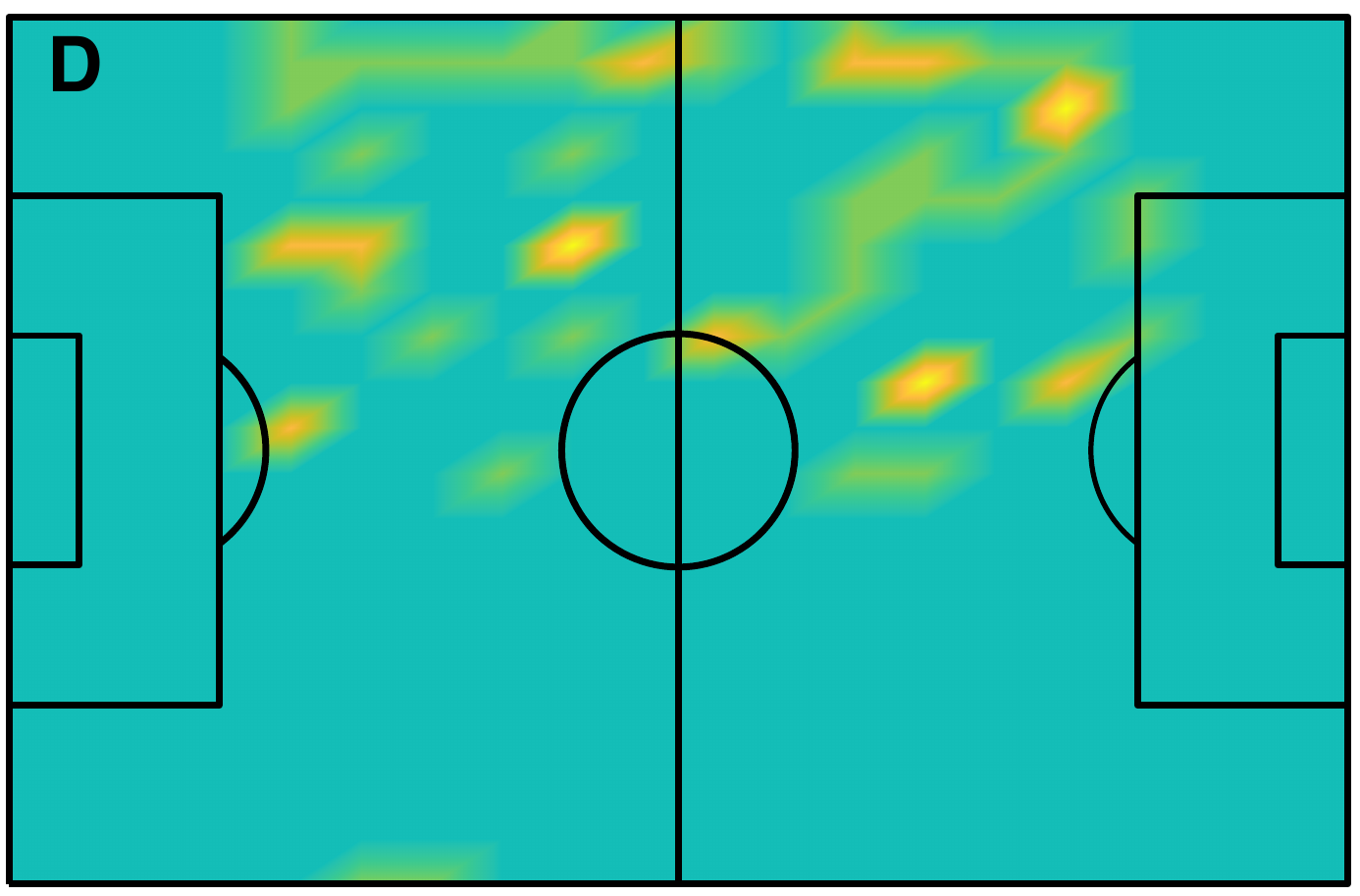}
\caption{Influence of the scale of the pitch partition. All plots correspond to the same event datasets, with the only difference that the number of partitions of the pitch increases from A to D. Specifically, all partitions correspond to a $M \times M$ lattice with: (A) $M=4$, i.e. $P=16$ divisions; (B) $M=6$, i.e. $P=36$ divisions, (C) $M=10$, i.e. $P=100$ divisions and (D) $M=20$, i.e. $P=400$ divisions.  \label{f02}}
\end{figure}

One of the conclusions drawn from Fig. ~\ref{f01} is that there is a discrepancy between event and tracking heatmaps. This is something expected since they measure different things, however, sometimes, the event heatmap can be used as a proxy of the tracking heatmap. This is the case of a diversity of sports media who do not have access to the tracking datasets. The main limitation of this approach is that events lead to a discrete (i.e., opposite to continuous) representation of the player position. This issue is highlighted in Fig. ~\ref{f02}, where we represented the event heatmaps of a given player using different scales. In this case, the term "scale" refers to the number of partitions used to average the events of a player. Therefore, to elaborate Fig.~\ref{f02} we divided the pitch into $P=M\times M$ parts with P=16 (A), $P=36$ (B), $P=100$ (C) and $P=400$ (D) divisions. Next, we counted the number of events of a player inside each division and, finally, interpolated the value with adjacent cells to smooth the colour gradient. Note that the size of the grid leads to different heatmaps, despite the datasets are the same. Therefore, one of the crucial point when plotting event heatmaps is the election of the scale.

 %%%%%%%%%%%%%%%%%%% FIGURE 02
 %%%%%%%%%%%%%%%%%%% FIGURE 02

The mathematical representation of each heatmap is a $M \times M$ matrix that  contains the total number of events (or time, in the case of tracking heatmaps) a player has carried out inside each of the $P=M \times M$ partitions. We called  $\bf{E}$(i,j) the matrix corresponding to the event-based heatmap of player $i$  during match $j$. In the same way, the matrix $\bf{T}$(i,j) contains the times spent by player $i$ in match $j$ at each of the $P$ partitions of the pitch. Next, we calculated the 2-D correlation coefficient $C(E,T)$ between matrices $\bf{E}$(i,j) and $\bf{T}$(i,j):

\begin{equation}
C(E,T) = \frac{\sum_m\sum_n(E_{mn}- \bar{E})(T_{mn}- \bar{T})}{\sqrt{(\sum_m\sum_n(E_{mn}- \bar{E})^2)(\sum_m\sum_n(T_{mn}- \bar{T})^2)}} 
\end{equation}
where $\bar{E}$ and $\bar{T}$ are the average values of matrices $\bf{E(i,j)}$ and $\bf{T(i,j)}$. In this way, we obtained the correlation $C(E,T)$ of the event and tracking heatmaps of $N=1320$ player participations during the $L= 60$ matches.

\section*{Results}

\begin{figure}[t]
\centering
\includegraphics[width=15.5 cm]{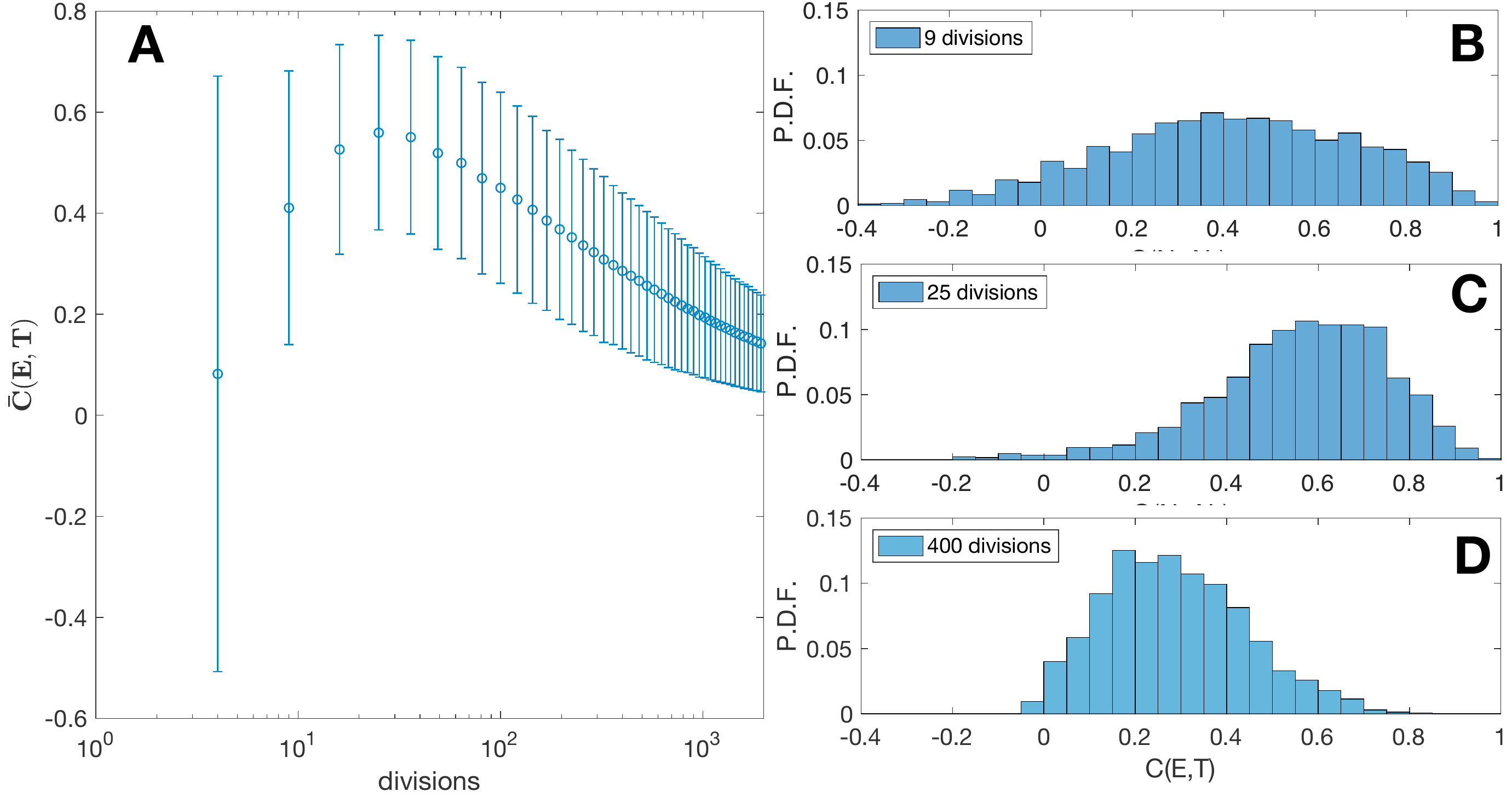}
\caption{Influence of the scale of the pitch partition on the correlation between event and tracking heatmaps. In (A), we plot the average correlation $\bar{C}(E,T)$ between $E(i,j)$ and $T(i,j)$ as a function of the number of divisions of the pitch. Error bars account for the corresponding standard deviation. Note the logarithmic scale in the X-axis. Plots (B-D) show the probability distribution functions for three particular partitions:  $P=9$ (B),  $P=25$ (C), which corresponds to the maximum value of  $\bar{C}(E,T)$,  and  $P=400$ (D).   \label{f03}}
\end{figure}

Figure~\ref{f03} shows the average correlation $\bar{C}(E,T)$ between the matrices $\bf{E}$ and $\bf{T}$ associated, respectively, to the event and tracking datasets. Average correlations have been calculated at different scales. Specifically, we have considered $P=M \times M$ divisions with $M=2,3,4, ..., 50$ (i.e., $P=4,9,16, ..., 2500$ divisions). Fig. ~\ref{f03}(A) shows that the average correlation has a maximum value at $M=5$ ($P=25$ partitions), indicating that this is the scale at which the event and tracking heatmaps correlate the most. However, two facts are indicating that both approaches are far from being equivalent. First, the average correlation for the best partition is $\bar{C}_{max}=0.56$, which is a rather low value. Second, the error bars, consisting of the standard deviation of the correlations of all players, are quite large, revealing that there is high heterogeneity in the correlation values. This is what we observe in Figs.~\ref{f03}(B-C-D), where we plot the probability distribution functions (PDFs) of the correlations  $C(E,T)$ at three different scales: $P=9$ (B), $P=25$ (i.e., the best partition) and $P=400$ (D). Interestingly, we can see that the distributions are quite wide. Furthermore, negative values of the correlation coexist with very high ones. Even in the best partition (C) we reported cases of negative correlations between the event and tracking heatmaps. Although, in this case, the peak of the distribution is closer to one, just a few cases are higher than 0.9. When the pitch is divided into further divisions (D), the maximum gets closer to zero and correlations close to the unity vanish.

\begin{figure}[t]
\centering
\includegraphics[width=9 cm]{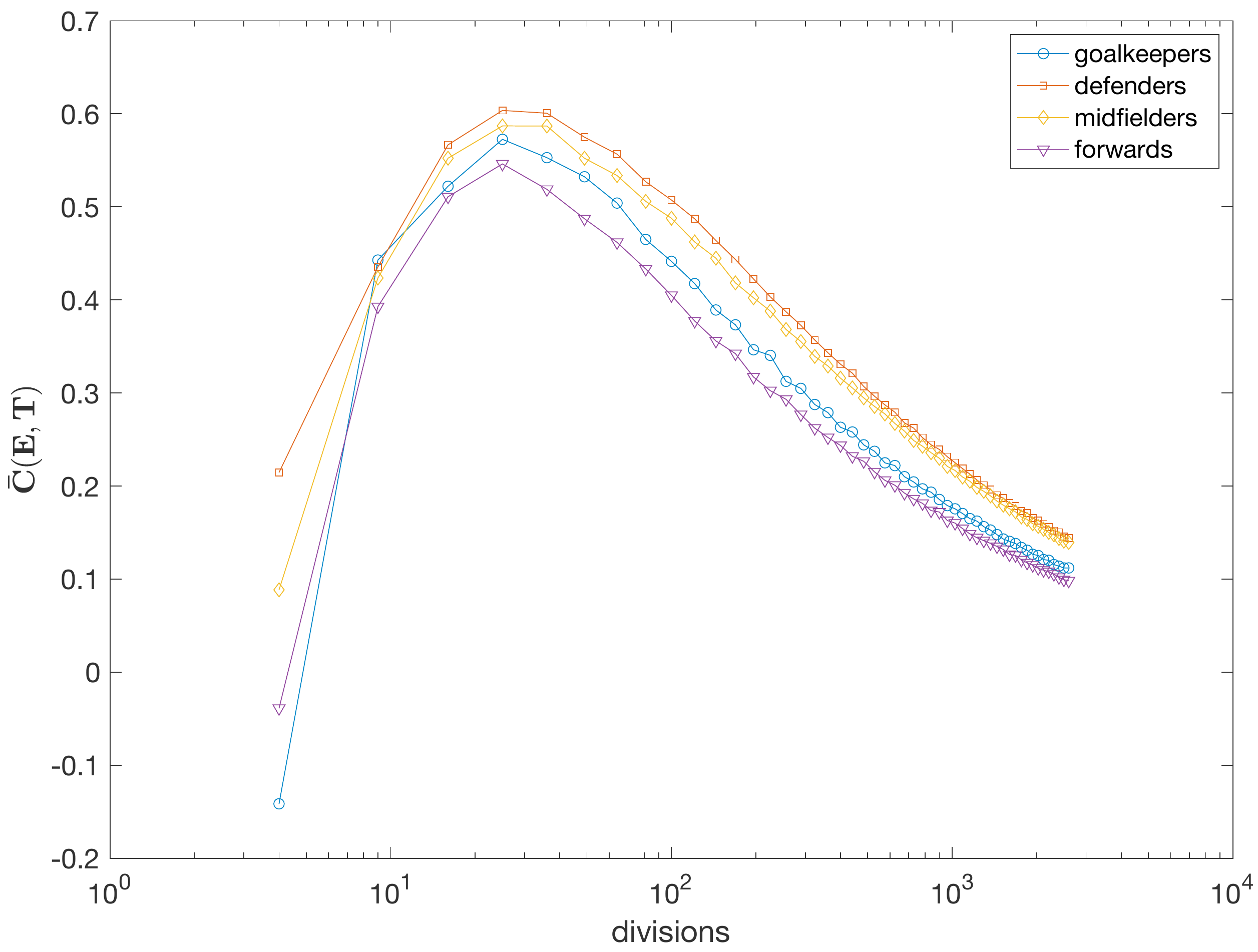}
\caption{Influence of the player position on the correlation between event and tracking heatmaps. We plot the average correlation between $E(i,j)$ and $T(i,j)$ as a function of the number of divisions of the pitch. Four different categories were considered: goalkeepers, defenders, midfielders and forwards (see figure legend). Note the logarithmic scale in the X-axis.  \label{f04}}
\end{figure}
\begin{figure}[h!]
\centering
\includegraphics[width=9.3 cm]{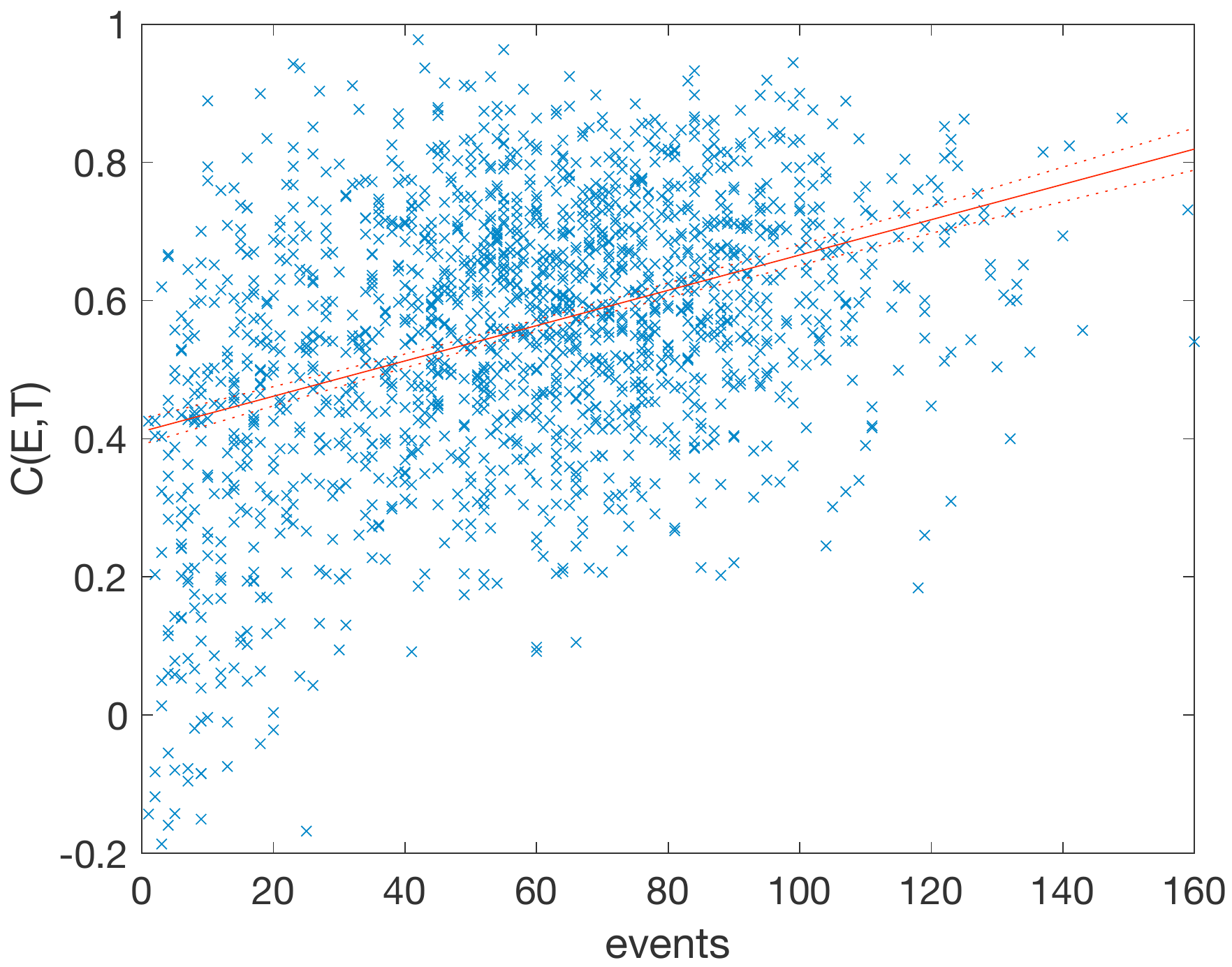}
\caption{Correlation between heatmaps $C(E,T)$ vs the number of events $n$ of a player.  Each point corresponds to a player participation. The red line is the linear fit, whose equation is $C(E,T)=0.0025551 n + 0.41045$. Dashed lines indicate the estimate of a $95\%$ prediction interval, with an r-squared of $0.154$. \label{f05}}
\end{figure}

As explained in Ref. \cite{konefal(2019)}, the player position strongly determines the movements and actions of players. For this reason, we decided to recalculate correlations at the different scales with a filter based on the player position. Four different categories were considered: (i) goalkeepers, (ii) defenders, (iii) midfielders and (iv) forwards. As we can see in Fig.~\ref{f04} all positions have a similar behaviour, with a peak around $P=25$ divisions. However, it is worth noting that certain positions have higher correlations than others. Specifically, defenders are the players with the highest correlation ($C^{def}_{max}=0.60$), followed, respectively, by midfielders $C^{mid}_{max}=0.59$, goalkeepers $C^{gk}_{max}=0.57$ and forwards $C^{fwd}_{max}=0.55$. This order is maintained for all partitions higher than $P=9$.

Finally, in Fig.~\ref{f05}, we plot the value of each particular player participation vs its corresponding number of events. The main objective was to understand whether the number of events could be related to the value of the correlation or, in other words, if a higher number of events could imply better correlations between event and tracking heatmaps. Fig.~\ref{f05} shows that negative correlations always occur when the total number of events associated with a player is low. However, the points' dispersion is rather high. A linear fit (red line) confirms these intuitions. We obtained the equation  $C(E,T)=0.0025551 n + 0.41045$, with $n$ being the number of events. Although the slope of the linear fit is positive with a p-value of $p=9.31 e^{-63}$ (dashed lines indicate the estimate of a $95\%$ prediction interval), the r-squared is 0.154. Therefore, the linear relation between the number of events and the correlation between the event and tracking heatmaps only explains the $15.4\%$ of the variability. In other words, there are other variables not considered in this linear model that influence the correlation between both heatmaps more drastically than the number of events.

\section*{Conclusions}

We investigated what is the relation between players' heatmaps when they are obtained using event or tracking datasets. Our results show that the correlation between both kinds of heatmaps is relatively low and, at the same time, there exists a high heterogeneity, with correlations ranging from negative values to values close to one. Correlations strongly depend on the division of the pitch made to construct the probability distribution of the events. We have seen that there is an optimal pitch partition ($P=36$ divisions) where both approaches correlate the most. In any case, even at the best partition, correlations show a high standard deviation. Correlations are slightly influenced by the number of events assigned to a player, however, this variable only explains around the $15\%$ of the variance. 

In accordance with recent studies relating the position of players with the events they perform \cite{konefal(2019)}, we observed that the average correlation is also affected by the player position. Defenders are the players whose event and tracking heatmaps overlap the most, while forwards are those with the lowest correlation. This behaviour may be motivated by the fact that defenders are the field players that have to maintain their positions the most while, on the contrary, forwards' movements have more freedom with the aim of creating unexpected situations for the opponents.

Finally, it is worth noting that the low correlation between both kinds of heatmaps is, somehow, expected since they measure different things. Event heatmaps indicate where a player has performed more actions, which is interesting information about player performance, probably with a more tactical influence. On the other hand, tracking events inform about the position of a player without adding information about how relevant his performance was. Therefore, we believe that both kinds of heatmaps can be complementary but, in any case, they should not be confused or treated as equals.

\section*{Acknowledgements}

Authors acknowledge Paco Seirul.lo (F.C. Barcelona), Xavier Busquets (E.S.A.D.E.) and Mat\'ias Conde (StatsPerform \& Big Data Sports) for fruitful conversations. J.M.B. is funded by MINECO (FIS2017-84151-P).  D.G. is funded by Comunidad de Madrid, Spain, through project PEJ-2018-AI/TIC-11183. 

\section*{Author contributions statement}

All authors participated in the conception of the article. R.L.dC. and R.R. provided the event and tracking datasets. D.G., D.R.A. and J.M.B. carried out the analysis. J.M.B. wrote the initial draft. All authors revised the manuscript together; all authors gave final approval for publication. 

\section*{Additional information}

\textbf{Competing interests:} Authors declare no competing interests. 


\begin{thebibliography}{}

\bibitem{clemente(2013)}
Clemente, F.M.; Couceiro, M.S.; Martins, F.M.L. Soccer teams behaviors: analysis of the team’s distribution in function to ball possession. {\em Research Journal of Applied Sciences, Engineering and Technology} {\bf 6}, 130--136 (2013).

\bibitem{moura(2017)}
Moura, F.A.; Santana, J.E.; Vieira, N.A.; Santiago, P.R.P.;Cunha, S.A. Analysis of soccer players' positional variability during the 2012 UEFA European Championship: A case study. {\em Journal of Human Kinetics}  {\bf 47}, 225 (2015).

\bibitem{bialkowski(2014)}
Bialkowski, A., Lucey, P., Carr, P.; Yue, Y., Sridharan, S. \& Matthews, I. Identifying team style in soccer using formations learned from spatiotemporal tracking data. In {\em 2014 IEEE international conference on data mining workshop}, 9--14 (2014).

\bibitem{machado(2017)}
Machado, V., Leite, R., Moura, F., Cunha, S., Sadlo, F. \& Comba, J.L. Visual soccer match analysis using spatiotemporal positions of players. {\em Computers \& Graphics} {\bf 68}, 84--95 (2014). 

\bibitem{gudmundsson(2017)}
Gudmundsson, J. \& Horton, M. Spatio-temporal analysis of team sports. {\em ACM Comput. Surv.} {\bf 50}, 22 (2017). 

\bibitem{suarez}
Luis Suárez heatmap. Available on line: https://www.dailymail.co.uk/sport/football/article-8629207 [accessed on 10th April 2021].

\bibitem{mediacoach}
Mediacoach. Available on line: https://www.mediacoach.es [accessed on 10th April 2021].

\bibitem{linke(2020)}
Linke, D., Link, D. \& Lames, M. Football-specific validity of TRACAB's optical video tracking systems. {\em PLoS ONE} {\bf 15}, e0230179 (2020).

\bibitem{felipe(2019)}
Felipe, J.L., Garcia-Unanue, J., Viejo-Romero, D., Navandar, A. \& S\'anchez-S\'anchez, J. Validation of a video-based performance analysis system (Mediacoach®) to analyze the physical demands during matches in LaLiga. {\em Sensors} {\bf 19}, 4113 (2019).

\bibitem{pons(2019)}
Pons, E., Garc\'ia-Calvo, T., Resta, R., Blanco, H., L\'opez Del Campo, R., Díaz Garc\'ia, J. \&  Pulido, J.J.  A comparison of a GPS device and a multi-camera video technology during official soccer matches: Agreement between systems. {\em PloS ONE}, {\bf 14}, e0220729 (2019).

\bibitem{opta}
Opta. Available on line: http://www.optasports.com [accessed on 10th April 2021].

\bibitem{optadescription}
Desctiption of Opta events. Available on line: https://www.statsperform.com/opta-event-definitions [accessed on 10th April 2021].

\bibitem{konefal(2019)}
Konefal, M., Chmura, P., Zając, T., Chmura, J., Kowalczuk, E. \& Andrzejewski, M.  A new approach to the analysis of pitch-positions in professional soccer. {\em Journal of Human Kinetics} {\bf 66}, 143 (2019).

\end{thebibliography}
\end{document}